%% file: ttc13-funnyqt-pn2sc.tex
\documentclass[submission]{eptcs}
\providecommand{\event}{Transformation Tool Contest 2013 (TTC'13)}
\def\titlerunning{Solving the Petri-Nets to Statecharts Transformation Case with FunnyQT}
\def\authorrunning{Tassilo Horn}

\usepackage[utf8]{inputenc}
\usepackage[T1]{fontenc}
\usepackage{hyperref}
\usepackage{paralist}
\usepackage{footmisc}

\makeatletter
\def\verbatim@font{\ttfamily\small}
\makeatother

\usepackage{color}
\usepackage{fancyvrb}
\input{pyg}

\title{Solving the Petri-Nets to Statecharts Transformation Case with FunnyQT}
\author{Tassilo Horn
  \email{horn@uni-koblenz.de}
  \institute{Institute for Software Technology, University Koblenz-Landau, Germany}}

\begin{document}

\maketitle

\begin{abstract}
  FunnyQT is a model querying and model transformation library for the
  functional Lisp-dialect Clojure providing a rich and efficient querying and
  transformation API.

  This paper describes the FunnyQT solution to the TTC 2013 Petri-Nets to
  Statcharts Transformation Case.  This solution has won the \emph{best overall
    solution award} and the \emph{best efficiency award} for this case.
\end{abstract}

\section{Introduction}
\label{sec:introduction}

\emph{FunnyQT}\footnote{The FunnyQT homepage:
  \url{https://github.com/jgralab/funnyqt}} is a new model querying and
transformation approach which is implemented as an API for the functional,
JVM-based Lisp-dialect Clojure.  It provides several sub-APIs for implementing
different kinds of queries and transformations.  For example, there is a
model-to-model transformation API, and there is an in-place transformation API
for writing programmed graph transformations.  FunnyQT currently supports EMF
and JGraLab models, and it can be extended to other modeling frameworks, too.

For solving the tasks of this transformation case\footnote{This FunnyQT
  solution is available at \url{https://github.com/tsdh/ttc-2013-pn2sc} and on
  SHARE (image
  \textsf{TTC13::Ubuntu12LTS\_TTC13::FunnyQT.vdi})\label{fn:github}}, FunnyQT's
model transformation API has been used for the initialization transformation,
while the reduction transformation has been tackled algorithmically using the
plain querying and model manipulation APIs.  This solution has won the
\emph{best overall solution award} and the \emph{best efficiency award} for
this case.

\section{The Initialization Transformation}
\label{sec:init-transformation}

The initialization transformation using FunnyQT's model transformation API is
shown in Listing~\ref{lst:init}.  This API provides an internal DSL
\cite{book:Fowler2010DSL} for model-to-model transformations similar to ATL
\cite{ATL05} or ETL \cite{booklet:epsilon}.

A transformation is declared with the \verb|deftransformation| macro.  It
receives the name of the transformation, i.e., \verb|initialize-statechart|, a
vector of input and output models, and arbitrarily many rules.  Here, the
argument vector declares that the transformation receives one single input
model \verb|pn| which is an EMF model, and it receives exactly one output model
\verb|sc| which is also an EMF model.  It could also receive many input and
output models, and those could belong to different modeling frameworks as well.

The transformation consists of two rules: \verb|place2basic-and-or|, and
\verb|transition2hyperedge|.  The former receives an input \verb|Place| and
creates an \verb|OR| and a \verb|Basic| in the output model.  It also sets the
new basic's name to the name of the place and assigns it as content of the new
\verb|OR|.  Finally, it sets the basic's \verb|rnext| and \verb|next|
references to the value of applying \verb|transition2hyperedge| to any
pre-transition or post-transition of the input place\footnote{\textsf{map}
  takes a function $f$ and a collection $c$.  It applies $f$ to the items of
  $c$ returning the sequence of results.}.  \verb|place2basic-and-or| is a
top-level rule meaning it's applied automatically to all matching elements
while the non-top-level rule \verb|transition2hyperedge| has to be called
explicitly from other rules.

\begin{figure}[h!]
  \input{lst/01}
  \label{lst:init}
  \caption{The initialization transformation}
\end{figure}

When a rule gets called and is applicable with respect to its declared
\verb|:from| type (and optional \verb|:when| constraint), it creates the
elements declared in \verb|:to| in the target model, and evaluates its body.
In case there is just one new element declared in \verb|:to|, it returns just
that.  If there are many new elements, it returns them as a vector in their
declaration order.  Furthermore, a traceability mapping is created from the
source element to the rule's return value.  If a rule gets called multiple
times for a single element, the second and all following calls just return the
result of the first invocation.

\section{The Reduction Transformation}
\label{sec:reduction-transformation}

The reduction transformation is implemented algorithmically based on FunnyQT's
querying and model manipulation APIs.  It consists of four rules (functions):
\begin{compactenum}
\item The AND-rule as discussed in the case description \cite{pn2sccasedesc},
\item the OR-rule as discussed in the case description,
\item an additional, extension rule assigning hyperedges to the nearest
  \verb|Compound| state containing all their predecessor and successor
  \verb|Basic| states,
\item and a rule creating a \verb|Statechart| with an \verb|AND| top-state if
  the reduction could be completed successfully.
\end{compactenum}

In this section, only the AND- and OR-rule are discussed.  The main reduction
function simply applies them as long as possible, then invokes the hyperedge
assignment rule followed by the statechart creating rule.  The complete
reduction transformation is printed in Appendix~\ref{sec:reduction-code}.

\paragraph{Reduction Helper Functions.}
\label{sec:reduct-help-functions}

Before discussing the rules, some helper functions need to be introduced.
Those are \verb|pret| and \verb|postt| returning the sets of
pre-/post-transitions for a given place.  Likewise, \verb|prep| and
\verb|postp| return the sets of pre-/post-places for a given transition.

\paragraph{The AND Rule.}
\label{sec:and-rule}

The AND rule is depicted in Listing~\ref{lst:and-rule}.  In contrast to the
Figure~2 in the case description \cite{pn2sccasedesc}, it doesn't delete all
places $q_1$ to $q_n$ to create a new place $p$, but instead it reuses $q_1$ as
$p$ and deletes only $q_2$ to $q_n$, which is consistent with Louis Rose's EOL
solution.

The rule function receives the source Petri-net model \verb|pn|, the target
statechart model \verb|sc|, either the function \verb|prep| or \verb|postp| as
\verb|prep-or-postp|, and the traceability map \verb|place2or| gathered from
the initialization transformation mapping input places to output \verb|OR|
states wrapped in a Clojure atom\footnote{All Clojure data structures are
  immutable.  An atom is a mutable reference to some immutable data structure
  that can be swapped atomically.  This is important here in order to update
  the traceability mapping when the AND-rule matches.}.

\begin{figure}[h!]
  \input{lst/02}
  \label{lst:and-rule}
  \caption{The AND rule}
\end{figure}

The rule iterates over all transitions\footnote{\textsf{seq} takes a collection
  and returns a sequential view on it or \textsf{nil} if the collection is
  empty.  Therefore, it is the canonical non-emptiness check in Clojure.} in
the petri-net model \verb|pn| using a local tail-recursion (\verb|loop| and
\verb|recur|).  Lines 5 to 9 check the preconditions of the rule: If the
transition \verb|t| has more than one pre-/post-place, all of them must have
the same set of pre- and post-transitions.  If that's the case, a new
\verb|AND| and a new \verb|OR| state is created.  The new \verb|AND| contains
all existing \verb|OR| states being the pre- or post-places of the transition
\verb|t|, and the new \verb|OR| contains the new \verb|AND|.  Furthermore, the
traceability map atom is updated in line 13 so that the first pre-/post-place
\verb|p| now maps to the new \verb|OR|.  Lastly, all other pre-/post-places are
deleted\footnote{\textsf{doseq} is equivalent to Java's extended \textsf{for}
  loop.}.

\paragraph{The OR Rule.}
\label{sec:or-rule}

The OR rule is depicted in Listing~\ref{lst:or-rule}.  In contrast to the case
description, it doesn't delete the places (or corresponding OR states) $q$ and
$r$ to create a new place (or corresponding OR state) $p$, but instead it
reuses $q$ as $p$ and only deletes $r$.

The \verb|or-rule| gets the Petri-net model \verb|pn|, the statechart model
\verb|sc|, and the traceability map atom \verb|place2or|.

\begin{figure}[h!t]
  \input{lst/03}
  \label{lst:or-rule}
  \caption{The OR rule}
\end{figure}

Like the AND-rule, it iterates all transitions in the petri-net model.  Lines 5
to 9 check the preconditions of the rule: If the transition \verb|t| has
exactly one pre-place \verb|q| and one post-place \verb|r|, and if \verb|q| and
\verb|r| are identical or \verb|q| and \verb|r| are not connected by other
transitions, then the rule matches.  In that case, the \verb|OR| corresponding
to \verb|r| is merged with the \verb|OR| corresponding to \verb|q|, and the
transition \verb|t| is deleted.

\paragraph{Extensions.}
\label{sec:extensions}

\begin{sloppypar}
  Two extensions were implemented for this task.  Firstly, there is an
  additional rule \verb|assign-hyperedges| that assigns each hyperedge to the
  nearest compound state which contains all basic states connected by the
  hyperedge.  Secondly, a validation
  tool\footnote{\url{https://github.com/tsdh/ttc-2013-pn2sc-validation}} has
  been implemented that uses FunnyQT to check result statechart models against
  their expected outcome in terms of a very detailed unit test suite.
\end{sloppypar}
\section{Evaluation}
\label{sec:evaluation}

In this section, the solution is evaluated according to the evaluation criteria
listed in the case description~\cite{pn2sccasedesc}.

\paragraph{Transformation correctness.}

The validation project that has been implemented as an extension to this case
allows for testing the result statechart models.  For the main test cases,
every important aspect of the result models including the containment hierarchy
and the predecessors and successors of hyperedges are checked, and for the
performance test cases, only the number of instances of every metamodel class
is checked.  All tests pass for the result models of this solution.  Similarly,
all tests pass for the result models created by the reference GrGen.NET
solution.

The validation project has also been tested with intentionally slightly wrong
models, e.g., some \verb|next| link is missing at some hyperedge, there's some
additional element, or an element is contained by the wrong \verb|Compound|
state.  In all those cases, an assertion of the validation project failed.  So
there's a high confidence that if the result models pass the tests, the
transformation producing them is correct.

\paragraph{Transformation performance.}

This FunnyQT solution is by far the most efficient of all submitted solutions,
especially for large models, so it has won the \emph{best efficiency award} for
this case.  For the performance test models \verb|sp5000| and \verb|sp10000|
the complete transformation (initialization and reduction) takes about one and
two seconds, which is about as fast as the second fastest solution.  But with
the \verb|sp40000|/\verb|sp200000| model, the FunnyQT solution is already
three/twelve times faster than the second fastest solution, taking 11 and 114
seconds, respectively.

\paragraph{Transformation understandability.}

Although the solution requires some understanding of Clojure, it shouldn't be
hard to get a grasp on it.  The initialization transformation uses a FunnyQT
facility allowing to specify typical model transformations with a syntax and
semantics similar to ATL or ETL, so people knowing these languages should feel
right at home.

The reduction transformation is a bit more complex, but the application
conditions of the rules and the actions that are performed are taken quite
literally from the case description with the exception that some elements are
preserved and merged instead of replaced.

One important aspect with respect to understandability is also the fact that
the transformations are very concise.  In total, the initialization and the
reduction transformation are only 96 lines of code.

\paragraph{Bonus criteria.}

The bonus tasks dealing with \emph{verification}, \emph{simulation support},
\emph{change propagation}, and \emph{reversing the transformation} haven't been
tackled.

Proper \emph{debugging support} is also not yet ready for prime-time in the
Clojure world.  There are some attempts at debuggers allowing to set
breakpoints and examine the lexical extent around the breakpoint, but those are
not too usable right now.  Another difficulty with functional languages
involving some kind of laziness is that errors might be signaled at a location
very different to where the bug is actually manifested in the source code.
Nevertheless, FunnyQT has rather good model visualization tools that have been
used while programming the reduction rules in order to visualize the matching
elements when a rule has been applicable.

\bibliographystyle{eptcs}
\bibliography{ttc13-funnyqt-pn2sc}

\appendix
\newpage

\section{The complete Initialization Transformation}
\label{sec:init-code}

\input{lst/complete-init}

\section{The complete Reduction Transformation}
\label{sec:reduction-code}

\input{lst/complete-reduction}

\end{document}

%% file: lst/01.tex
\begin{Verbatim}[commandchars=\\\{\},numbers=left,firstnumber=1,stepnumber=1,fontsize=\footnotesize]
\PY{p}{(}\PY{n+nf}{deftransformation} \PY{n+nv}{initialize\PYZhy{}statechart} \PY{p}{[}\PY{p}{[}\PY{n+nv}{pn} \PY{l+s+ss}{:emf}\PY{p}{]} \PY{p}{[}\PY{n+nv}{sc} \PY{l+s+ss}{:emf}\PY{p}{]}\PY{p}{]}
  \PY{p}{(}\PY{o}{\PYZca{}}\PY{l+s+ss}{:top} \PY{n+nv}{place2basic\PYZhy{}and\PYZhy{}or} \PY{p}{[}\PY{n+nv}{p}\PY{p}{]}
         \PY{l+s+ss}{:from} \PY{l+s+ss}{\PYZsq{}Place}
         \PY{l+s+ss}{:to} \PY{p}{[}\PY{n+nv}{o} \PY{l+s+ss}{\PYZsq{}OR}, \PY{n+nv}{b} \PY{l+s+ss}{\PYZsq{}Basic}\PY{p}{]}
         \PY{p}{(}\PY{n+nf}{eset!} \PY{n+nv}{b} \PY{l+s+ss}{:name} \PY{p}{(}\PY{n+nf}{eget} \PY{n+nv}{p} \PY{l+s+ss}{:name}\PY{p}{)}\PY{p}{)}
         \PY{p}{(}\PY{n+nf}{eset!} \PY{n+nv}{b} \PY{l+s+ss}{:rcontains} \PY{n+nv}{o}\PY{p}{)}
         \PY{p}{(}\PY{n+nf}{eset!} \PY{n+nv}{b} \PY{l+s+ss}{:rnext} \PY{p}{(}\PY{n+nb}{map }\PY{n+nv}{transition2hyperedge} \PY{p}{(}\PY{n+nf}{eget} \PY{n+nv}{p} \PY{l+s+ss}{:pret}\PY{p}{)}\PY{p}{)}\PY{p}{)}
         \PY{p}{(}\PY{n+nf}{eset!} \PY{n+nv}{b} \PY{l+s+ss}{:next}  \PY{p}{(}\PY{n+nb}{map }\PY{n+nv}{transition2hyperedge} \PY{p}{(}\PY{n+nf}{eget} \PY{n+nv}{p} \PY{l+s+ss}{:postt}\PY{p}{)}\PY{p}{)}\PY{p}{)}\PY{p}{)}
  \PY{p}{(}\PY{n+nf}{transition2hyperedge} \PY{p}{[}\PY{n+nv}{t}\PY{p}{]}
         \PY{l+s+ss}{:from} \PY{l+s+ss}{\PYZsq{}Transition}
         \PY{l+s+ss}{:to} \PY{p}{[}\PY{n+nv}{he} \PY{l+s+ss}{\PYZsq{}HyperEdge}\PY{p}{]}
         \PY{p}{(}\PY{n+nf}{eset!} \PY{n+nv}{he} \PY{l+s+ss}{:name} \PY{p}{(}\PY{n+nf}{eget} \PY{n+nv}{t} \PY{l+s+ss}{:name}\PY{p}{)}\PY{p}{)}\PY{p}{)}\PY{p}{)}
\end{Verbatim}

%% file: lst/02.tex
\begin{Verbatim}[commandchars=\\\{\},numbers=left,firstnumber=1,stepnumber=1,fontsize=\footnotesize]
\PY{p}{(}\PY{k+kd}{defn }\PY{n+nv}{and\PYZhy{}rule} \PY{p}{[}\PY{n+nv}{pn} \PY{n+nv}{sc} \PY{n+nv}{prep\PYZhy{}or\PYZhy{}postp} \PY{n+nv}{place2or}\PY{p}{]}
  \PY{p}{(}\PY{k}{loop }\PY{p}{[}\PY{n+nv}{ts} \PY{p}{(}\PY{n+nf}{eallobjects} \PY{n+nv}{pn} \PY{l+s+ss}{\PYZsq{}Transition}\PY{p}{)}, \PY{n+nv}{applied} \PY{n+nv}{false}\PY{p}{]}
    \PY{p}{(}\PY{k}{if }\PY{p}{(}\PY{n+nb}{seq }\PY{n+nv}{ts}\PY{p}{)}
      \PY{p}{(}\PY{k}{let }\PY{p}{[}\PY{n+nv}{t} \PY{p}{(}\PY{n+nb}{first }\PY{n+nv}{ts}\PY{p}{)}, \PY{n+nv}{preps\PYZhy{}or\PYZhy{}postps} \PY{p}{(}\PY{n+nf}{prep\PYZhy{}or\PYZhy{}postp} \PY{n+nv}{t}\PY{p}{)}\PY{p}{]}
        \PY{p}{(}\PY{k}{if }\PY{p}{(}\PY{n+nb}{\PYZgt{} }\PY{p}{(}\PY{n+nb}{count }\PY{n+nv}{preps\PYZhy{}or\PYZhy{}postps}\PY{p}{)} \PY{l+m+mi}{1}\PY{p}{)}
          \PY{p}{(}\PY{k}{let }\PY{p}{[}\PY{n+nv}{p} \PY{p}{(}\PY{n+nb}{first }\PY{n+nv}{preps\PYZhy{}or\PYZhy{}postps}\PY{p}{)}, \PY{n+nv}{prets} \PY{p}{(}\PY{n+nf}{pret} \PY{n+nv}{p}\PY{p}{)}, \PY{n+nv}{postts} \PY{p}{(}\PY{n+nf}{postt} \PY{n+nv}{p}\PY{p}{)}\PY{p}{]}
            \PY{p}{(}\PY{k}{if }\PY{p}{(}\PY{n+nf}{forall?} \PY{o}{\PYZsh{}}\PY{p}{(}\PY{n+nb}{and }\PY{p}{(}\PY{n+nb}{= }\PY{n+nv}{prets}  \PY{p}{(}\PY{n+nf}{pret} \PY{n+nv}{\PYZpc{}}\PY{p}{)}\PY{p}{)}
                               \PY{p}{(}\PY{n+nb}{= }\PY{n+nv}{postts} \PY{p}{(}\PY{n+nf}{postt} \PY{n+nv}{\PYZpc{}}\PY{p}{)}\PY{p}{)}\PY{p}{)}
                         \PY{p}{(}\PY{n+nb}{rest }\PY{n+nv}{preps\PYZhy{}or\PYZhy{}postps}\PY{p}{)}\PY{p}{)}
              \PY{p}{(}\PY{k}{let }\PY{p}{[}\PY{n+nv}{new\PYZhy{}or}  \PY{p}{(}\PY{n+nf}{ecreate!} \PY{n+nv}{sc} \PY{l+s+ss}{\PYZsq{}OR}\PY{p}{)}, \PY{n+nv}{new\PYZhy{}and} \PY{p}{(}\PY{n+nf}{ecreate!} \PY{n+nv}{sc} \PY{l+s+ss}{\PYZsq{}AND}\PY{p}{)}\PY{p}{]}
                \PY{p}{(}\PY{n+nf}{eset!} \PY{n+nv}{new\PYZhy{}and} \PY{l+s+ss}{:contains} \PY{p}{(}\PY{n+nf}{mapv} \PY{o}{@}\PY{n+nv}{place2or} \PY{n+nv}{preps\PYZhy{}or\PYZhy{}postps}\PY{p}{)}\PY{p}{)}
                \PY{p}{(}\PY{n+nf}{eadd!} \PY{n+nv}{new\PYZhy{}or}  \PY{l+s+ss}{:contains} \PY{n+nv}{new\PYZhy{}and}\PY{p}{)}
                \PY{p}{(}\PY{n+nf}{swap!} \PY{n+nv}{place2or} \PY{n+nb}{assoc }\PY{n+nv}{p} \PY{n+nv}{new\PYZhy{}or}\PY{p}{)}
                \PY{p}{(}\PY{n+nb}{doseq }\PY{p}{[}\PY{n+nv}{op} \PY{p}{(}\PY{n+nb}{rest }\PY{n+nv}{preps\PYZhy{}or\PYZhy{}postps}\PY{p}{)}\PY{p}{]}
                  \PY{p}{(}\PY{n+nf}{edelete!} \PY{n+nv}{op}\PY{p}{)}\PY{p}{)}
                \PY{p}{(}\PY{n+nf}{recur} \PY{p}{(}\PY{n+nb}{rest }\PY{n+nv}{ts}\PY{p}{)} \PY{n+nv}{true}\PY{p}{)}\PY{p}{)}
              \PY{p}{(}\PY{n+nf}{recur} \PY{p}{(}\PY{n+nb}{rest }\PY{n+nv}{ts}\PY{p}{)} \PY{n+nv}{applied}\PY{p}{)}\PY{p}{)}\PY{p}{)}
          \PY{p}{(}\PY{n+nf}{recur} \PY{p}{(}\PY{n+nb}{rest }\PY{n+nv}{ts}\PY{p}{)} \PY{n+nv}{applied}\PY{p}{)}\PY{p}{)}\PY{p}{)}
      \PY{n+nv}{applied}\PY{p}{)}\PY{p}{)}\PY{p}{)}
\end{Verbatim}

%% file: lst/03.tex
\begin{Verbatim}[commandchars=\\\{\},numbers=left,firstnumber=1,stepnumber=1,fontsize=\footnotesize]
\PY{p}{(}\PY{k+kd}{defn }\PY{n+nv}{or\PYZhy{}rule} \PY{p}{[}\PY{n+nv}{pn} \PY{n+nv}{sc} \PY{n+nv}{place2or}\PY{p}{]}
  \PY{p}{(}\PY{k}{loop }\PY{p}{[}\PY{n+nv}{ts} \PY{p}{(}\PY{n+nf}{vec} \PY{p}{(}\PY{n+nf}{eallobjects} \PY{n+nv}{pn} \PY{l+s+ss}{\PYZsq{}Transition}\PY{p}{)}\PY{p}{)}, \PY{n+nv}{applied} \PY{n+nv}{false}\PY{p}{]}
    \PY{p}{(}\PY{k}{if }\PY{p}{(}\PY{n+nb}{seq }\PY{n+nv}{ts}\PY{p}{)}
      \PY{p}{(}\PY{k}{let }\PY{p}{[}\PY{n+nv}{t} \PY{p}{(}\PY{n+nb}{first }\PY{n+nv}{ts}\PY{p}{)}, \PY{n+nv}{preps} \PY{p}{(}\PY{n+nf}{prep} \PY{n+nv}{t}\PY{p}{)}, \PY{n+nv}{postps} \PY{p}{(}\PY{n+nf}{postp} \PY{n+nv}{t}\PY{p}{)}\PY{p}{]}
        \PY{p}{(}\PY{k}{if }\PY{p}{(}\PY{n+nb}{= }\PY{l+m+mi}{1} \PY{p}{(}\PY{n+nb}{count }\PY{n+nv}{preps}\PY{p}{)} \PY{p}{(}\PY{n+nb}{count }\PY{n+nv}{postps}\PY{p}{)}\PY{p}{)}
          \PY{p}{(}\PY{k}{let }\PY{p}{[}\PY{n+nv}{q} \PY{p}{(}\PY{n+nb}{first }\PY{n+nv}{preps}\PY{p}{)}, \PY{n+nv}{r} \PY{p}{(}\PY{n+nb}{first }\PY{n+nv}{postps}\PY{p}{)}\PY{p}{]}
            \PY{p}{(}\PY{k}{if }\PY{p}{(}\PY{n+nb}{or }\PY{p}{(}\PY{n+nb}{identical? }\PY{n+nv}{q} \PY{n+nv}{r}\PY{p}{)}
                    \PY{p}{(}\PY{n+nb}{and }\PY{p}{(}\PY{n+nb}{not }\PY{p}{(}\PY{n+nf}{member?} \PY{n+nv}{r} \PY{p}{(}\PY{n+nf}{adjs} \PY{n+nv}{q} \PY{l+s+ss}{:pret} \PY{l+s+ss}{:postp}\PY{p}{)}\PY{p}{)}\PY{p}{)}
                         \PY{p}{(}\PY{n+nb}{not }\PY{p}{(}\PY{n+nf}{member?} \PY{n+nv}{r} \PY{p}{(}\PY{n+nf}{adjs} \PY{n+nv}{q} \PY{l+s+ss}{:postt} \PY{l+s+ss}{:prep}\PY{p}{)}\PY{p}{)}\PY{p}{)}\PY{p}{)}\PY{p}{)}
              \PY{p}{(}\PY{k}{let }\PY{p}{[}\PY{n+nv}{merger} \PY{p}{(}\PY{o}{@}\PY{n+nv}{place2or} \PY{n+nv}{q}\PY{p}{)}, \PY{n+nv}{mergee} \PY{p}{(}\PY{o}{@}\PY{n+nv}{place2or} \PY{n+nv}{r}\PY{p}{)}\PY{p}{]}
                \PY{p}{(}\PY{n+nb}{when\PYZhy{}not }\PY{p}{(}\PY{n+nb}{identical? }\PY{n+nv}{q} \PY{n+nv}{r}\PY{p}{)}
                  \PY{p}{(}\PY{n+nf}{eaddall!} \PY{n+nv}{q} \PY{l+s+ss}{:pret}  \PY{p}{(}\PY{n+nf}{eget\PYZhy{}raw} \PY{n+nv}{r} \PY{l+s+ss}{:pret}\PY{p}{)}\PY{p}{)}
                  \PY{p}{(}\PY{n+nf}{eaddall!} \PY{n+nv}{q} \PY{l+s+ss}{:postt} \PY{p}{(}\PY{n+nf}{eget\PYZhy{}raw} \PY{n+nv}{r} \PY{l+s+ss}{:postt}\PY{p}{)}\PY{p}{)}
                  \PY{p}{(}\PY{n+nf}{edelete!} \PY{n+nv}{r}\PY{p}{)}
                  \PY{p}{(}\PY{n+nf}{eaddall!} \PY{n+nv}{merger} \PY{l+s+ss}{:contains} \PY{p}{(}\PY{n+nf}{eget\PYZhy{}raw} \PY{n+nv}{mergee} \PY{l+s+ss}{:contains}\PY{p}{)}\PY{p}{)}
                  \PY{p}{(}\PY{n+nf}{edelete!} \PY{n+nv}{mergee}\PY{p}{)}\PY{p}{)}
                \PY{p}{(}\PY{n+nf}{edelete!} \PY{n+nv}{t}\PY{p}{)}
                \PY{p}{(}\PY{n+nf}{recur} \PY{p}{(}\PY{n+nb}{rest }\PY{n+nv}{ts}\PY{p}{)} \PY{n+nv}{true}\PY{p}{)}\PY{p}{)}
              \PY{p}{(}\PY{n+nf}{recur} \PY{p}{(}\PY{n+nb}{rest }\PY{n+nv}{ts}\PY{p}{)} \PY{n+nv}{applied}\PY{p}{)}\PY{p}{)}\PY{p}{)}
          \PY{p}{(}\PY{n+nf}{recur} \PY{p}{(}\PY{n+nb}{rest }\PY{n+nv}{ts}\PY{p}{)} \PY{n+nv}{applied}\PY{p}{)}\PY{p}{)}\PY{p}{)}
      \PY{n+nv}{applied}\PY{p}{)}\PY{p}{)}\PY{p}{)}
\end{Verbatim}

%% file: lst/complete-init.tex
\begin{Verbatim}[commandchars=\\\{\},numbers=left,firstnumber=1,stepnumber=1,fontsize=\footnotesize]
\PY{p}{(}\PY{n+nf}{deftransformation} \PY{n+nv}{initialize\PYZhy{}statechart} \PY{p}{[}\PY{p}{[}\PY{n+nv}{pn} \PY{l+s+ss}{:emf}\PY{p}{]} \PY{p}{[}\PY{n+nv}{sc} \PY{l+s+ss}{:emf}\PY{p}{]}\PY{p}{]}
  \PY{p}{(}\PY{o}{\PYZca{}}\PY{l+s+ss}{:top} \PY{n+nv}{place2basic\PYZhy{}and\PYZhy{}or} \PY{p}{[}\PY{n+nv}{p}\PY{p}{]}
         \PY{l+s+ss}{:from} \PY{l+s+ss}{\PYZsq{}Place}
         \PY{l+s+ss}{:to} \PY{p}{[}\PY{n+nv}{o} \PY{l+s+ss}{\PYZsq{}OR}, \PY{n+nv}{b} \PY{l+s+ss}{\PYZsq{}Basic}\PY{p}{]}
         \PY{p}{(}\PY{n+nf}{eset!} \PY{n+nv}{b} \PY{l+s+ss}{:name} \PY{p}{(}\PY{n+nf}{eget} \PY{n+nv}{p} \PY{l+s+ss}{:name}\PY{p}{)}\PY{p}{)}
         \PY{p}{(}\PY{n+nf}{eset!} \PY{n+nv}{b} \PY{l+s+ss}{:rcontains} \PY{n+nv}{o}\PY{p}{)}
         \PY{p}{(}\PY{n+nf}{eset!} \PY{n+nv}{b} \PY{l+s+ss}{:rnext} \PY{p}{(}\PY{n+nb}{map }\PY{n+nv}{transition2hyperedge} \PY{p}{(}\PY{n+nf}{eget} \PY{n+nv}{p} \PY{l+s+ss}{:pret}\PY{p}{)}\PY{p}{)}\PY{p}{)}
         \PY{p}{(}\PY{n+nf}{eset!} \PY{n+nv}{b} \PY{l+s+ss}{:next}  \PY{p}{(}\PY{n+nb}{map }\PY{n+nv}{transition2hyperedge} \PY{p}{(}\PY{n+nf}{eget} \PY{n+nv}{p} \PY{l+s+ss}{:postt}\PY{p}{)}\PY{p}{)}\PY{p}{)}\PY{p}{)}
  \PY{p}{(}\PY{n+nf}{transition2hyperedge} \PY{p}{[}\PY{n+nv}{t}\PY{p}{]}
         \PY{l+s+ss}{:from} \PY{l+s+ss}{\PYZsq{}Transition}
         \PY{l+s+ss}{:to} \PY{p}{[}\PY{n+nv}{he} \PY{l+s+ss}{\PYZsq{}HyperEdge}\PY{p}{]}
         \PY{p}{(}\PY{n+nf}{eset!} \PY{n+nv}{he} \PY{l+s+ss}{:name} \PY{p}{(}\PY{n+nf}{eget} \PY{n+nv}{t} \PY{l+s+ss}{:name}\PY{p}{)}\PY{p}{)}\PY{p}{)}\PY{p}{)}

\PY{p}{(}\PY{k+kd}{defn }\PY{n+nv}{init\PYZhy{}statechart} \PY{p}{[}\PY{n+nv}{pn}\PY{p}{]}
  \PY{p}{(}\PY{k}{let }\PY{p}{[}\PY{n+nv}{sc} \PY{p}{(}\PY{n+nf}{new\PYZhy{}model}\PY{p}{)}
        \PY{n+nv}{trace} \PY{p}{(}\PY{n+nf}{initialize\PYZhy{}statechart} \PY{n+nv}{pn} \PY{n+nv}{sc}\PY{p}{)}\PY{p}{]}
    \PY{p}{[}\PY{n+nv}{sc}
     \PY{p}{(}\PY{n+nb}{apply }\PY{n+nb}{hash\PYZhy{}map }\PY{p}{(}\PY{n+nb}{mapcat }\PY{p}{(}\PY{k}{fn }\PY{p}{[}\PY{p}{[}\PY{n+nv}{p} \PY{p}{[}\PY{n+nv}{o} \PY{n+nv}{b}\PY{p}{]}\PY{p}{]}\PY{p}{]} \PY{p}{[}\PY{n+nv}{p} \PY{n+nv}{o}\PY{p}{]}\PY{p}{)}
                             \PY{p}{(}\PY{l+s+ss}{:place2basic\PYZhy{}and\PYZhy{}or} \PY{n+nv}{trace}\PY{p}{)}\PY{p}{)}\PY{p}{)}
     \PY{p}{(}\PY{n+nb}{apply }\PY{n+nb}{hash\PYZhy{}map }\PY{p}{(}\PY{n+nb}{mapcat }\PY{p}{(}\PY{k}{fn }\PY{p}{[}\PY{p}{[}\PY{n+nv}{p} \PY{p}{[}\PY{n+nv}{o} \PY{n+nv}{b}\PY{p}{]}\PY{p}{]}\PY{p}{]} \PY{p}{[}\PY{n+nv}{p} \PY{n+nv}{b}\PY{p}{]}\PY{p}{)}
                             \PY{p}{(}\PY{l+s+ss}{:place2basic\PYZhy{}and\PYZhy{}or} \PY{n+nv}{trace}\PY{p}{)}\PY{p}{)}\PY{p}{)}
     \PY{p}{(}\PY{l+s+ss}{:transition2hyperedge} \PY{n+nv}{trace}\PY{p}{)}\PY{p}{]}\PY{p}{)}\PY{p}{)}
\end{Verbatim}

%% file: lst/complete-reduction.tex
\begin{Verbatim}[commandchars=\\\{\},numbers=left,firstnumber=1,stepnumber=1,fontsize=\footnotesize]
\PY{p}{(}\PY{k+kd}{defn }\PY{n+nv}{refs\PYZhy{}as\PYZhy{}set} \PY{p}{[}\PY{n+nb}{ref }\PY{n+nv}{elem}\PY{p}{]}
  \PY{p}{(}\PY{n+nb}{set }\PY{p}{(}\PY{n+nf}{eget\PYZhy{}raw} \PY{n+nv}{elem} \PY{n+nv}{ref}\PY{p}{)}\PY{p}{)}\PY{p}{)}

\PY{p}{(}\PY{k}{def }\PY{n+nv}{postt} \PY{p}{(}\PY{n+nb}{partial }\PY{n+nv}{refs\PYZhy{}as\PYZhy{}set} \PY{l+s+ss}{:postt}\PY{p}{)}\PY{p}{)}
\PY{p}{(}\PY{k}{def }\PY{n+nv}{pret}  \PY{p}{(}\PY{n+nb}{partial }\PY{n+nv}{refs\PYZhy{}as\PYZhy{}set} \PY{l+s+ss}{:pret}\PY{p}{)}\PY{p}{)}
\PY{p}{(}\PY{k}{def }\PY{n+nv}{postp} \PY{p}{(}\PY{n+nb}{partial }\PY{n+nv}{refs\PYZhy{}as\PYZhy{}set} \PY{l+s+ss}{:postp}\PY{p}{)}\PY{p}{)}
\PY{p}{(}\PY{k}{def }\PY{n+nv}{prep}  \PY{p}{(}\PY{n+nb}{partial }\PY{n+nv}{refs\PYZhy{}as\PYZhy{}set} \PY{l+s+ss}{:prep}\PY{p}{)}\PY{p}{)}

\PY{p}{(}\PY{k+kd}{defn }\PY{n+nv}{and\PYZhy{}rule} \PY{p}{[}\PY{n+nv}{pn} \PY{n+nv}{sc} \PY{n+nv}{prep\PYZhy{}or\PYZhy{}postp} \PY{n+nv}{place2or}\PY{p}{]}
  \PY{p}{(}\PY{k}{loop }\PY{p}{[}\PY{n+nv}{ts} \PY{p}{(}\PY{n+nf}{eallobjects} \PY{n+nv}{pn} \PY{l+s+ss}{\PYZsq{}Transition}\PY{p}{)}, \PY{n+nv}{applied} \PY{n+nv}{false}\PY{p}{]}
    \PY{p}{(}\PY{k}{if }\PY{p}{(}\PY{n+nb}{seq }\PY{n+nv}{ts}\PY{p}{)}
      \PY{p}{(}\PY{k}{let }\PY{p}{[}\PY{n+nv}{t} \PY{p}{(}\PY{n+nb}{first }\PY{n+nv}{ts}\PY{p}{)}, \PY{n+nv}{preps\PYZhy{}or\PYZhy{}postps} \PY{p}{(}\PY{n+nf}{prep\PYZhy{}or\PYZhy{}postp} \PY{n+nv}{t}\PY{p}{)}\PY{p}{]}
        \PY{p}{(}\PY{k}{if }\PY{p}{(}\PY{n+nb}{\PYZgt{} }\PY{p}{(}\PY{n+nb}{count }\PY{n+nv}{preps\PYZhy{}or\PYZhy{}postps}\PY{p}{)} \PY{l+m+mi}{1}\PY{p}{)}
          \PY{p}{(}\PY{k}{let }\PY{p}{[}\PY{n+nv}{p} \PY{p}{(}\PY{n+nb}{first }\PY{n+nv}{preps\PYZhy{}or\PYZhy{}postps}\PY{p}{)}, \PY{n+nv}{prets} \PY{p}{(}\PY{n+nf}{pret} \PY{n+nv}{p}\PY{p}{)}, \PY{n+nv}{postts} \PY{p}{(}\PY{n+nf}{postt} \PY{n+nv}{p}\PY{p}{)}\PY{p}{]}
            \PY{p}{(}\PY{k}{if }\PY{p}{(}\PY{n+nf}{forall?} \PY{o}{\PYZsh{}}\PY{p}{(}\PY{n+nb}{and }\PY{p}{(}\PY{n+nb}{= }\PY{n+nv}{prets}  \PY{p}{(}\PY{n+nf}{pret} \PY{n+nv}{\PYZpc{}}\PY{p}{)}\PY{p}{)}
                               \PY{p}{(}\PY{n+nb}{= }\PY{n+nv}{postts} \PY{p}{(}\PY{n+nf}{postt} \PY{n+nv}{\PYZpc{}}\PY{p}{)}\PY{p}{)}\PY{p}{)}
                         \PY{p}{(}\PY{n+nb}{rest }\PY{n+nv}{preps\PYZhy{}or\PYZhy{}postps}\PY{p}{)}\PY{p}{)}
              \PY{p}{(}\PY{k}{let }\PY{p}{[}\PY{n+nv}{new\PYZhy{}or}  \PY{p}{(}\PY{n+nf}{ecreate!} \PY{n+nv}{sc} \PY{l+s+ss}{\PYZsq{}OR}\PY{p}{)}, \PY{n+nv}{new\PYZhy{}and} \PY{p}{(}\PY{n+nf}{ecreate!} \PY{n+nv}{sc} \PY{l+s+ss}{\PYZsq{}AND}\PY{p}{)}\PY{p}{]}
                \PY{p}{(}\PY{n+nf}{eset!} \PY{n+nv}{new\PYZhy{}and} \PY{l+s+ss}{:contains} \PY{p}{(}\PY{n+nf}{mapv} \PY{o}{@}\PY{n+nv}{place2or} \PY{n+nv}{preps\PYZhy{}or\PYZhy{}postps}\PY{p}{)}\PY{p}{)}
                \PY{p}{(}\PY{n+nf}{eadd!} \PY{n+nv}{new\PYZhy{}or}  \PY{l+s+ss}{:contains} \PY{n+nv}{new\PYZhy{}and}\PY{p}{)}
                \PY{p}{(}\PY{n+nf}{swap!} \PY{n+nv}{place2or} \PY{n+nb}{assoc }\PY{n+nv}{p} \PY{n+nv}{new\PYZhy{}or}\PY{p}{)}
                \PY{p}{(}\PY{n+nb}{doseq }\PY{p}{[}\PY{n+nv}{op} \PY{p}{(}\PY{n+nb}{rest }\PY{n+nv}{preps\PYZhy{}or\PYZhy{}postps}\PY{p}{)}\PY{p}{]}
                  \PY{p}{(}\PY{n+nf}{edelete!} \PY{n+nv}{op}\PY{p}{)}\PY{p}{)}
                \PY{p}{(}\PY{n+nf}{recur} \PY{p}{(}\PY{n+nb}{rest }\PY{n+nv}{ts}\PY{p}{)} \PY{n+nv}{true}\PY{p}{)}\PY{p}{)}
              \PY{p}{(}\PY{n+nf}{recur} \PY{p}{(}\PY{n+nb}{rest }\PY{n+nv}{ts}\PY{p}{)} \PY{n+nv}{applied}\PY{p}{)}\PY{p}{)}\PY{p}{)}
          \PY{p}{(}\PY{n+nf}{recur} \PY{p}{(}\PY{n+nb}{rest }\PY{n+nv}{ts}\PY{p}{)} \PY{n+nv}{applied}\PY{p}{)}\PY{p}{)}\PY{p}{)}
      \PY{n+nv}{applied}\PY{p}{)}\PY{p}{)}\PY{p}{)}

\PY{p}{(}\PY{k+kd}{defn }\PY{n+nv}{or\PYZhy{}rule} \PY{p}{[}\PY{n+nv}{pn} \PY{n+nv}{sc} \PY{n+nv}{place2or}\PY{p}{]}
  \PY{p}{(}\PY{k}{loop }\PY{p}{[}\PY{n+nv}{ts} \PY{p}{(}\PY{n+nf}{vec} \PY{p}{(}\PY{n+nf}{eallobjects} \PY{n+nv}{pn} \PY{l+s+ss}{\PYZsq{}Transition}\PY{p}{)}\PY{p}{)}, \PY{n+nv}{applied} \PY{n+nv}{false}\PY{p}{]}
    \PY{p}{(}\PY{k}{if }\PY{p}{(}\PY{n+nb}{seq }\PY{n+nv}{ts}\PY{p}{)}
      \PY{p}{(}\PY{k}{let }\PY{p}{[}\PY{n+nv}{t} \PY{p}{(}\PY{n+nb}{first }\PY{n+nv}{ts}\PY{p}{)}, \PY{n+nv}{preps} \PY{p}{(}\PY{n+nf}{prep} \PY{n+nv}{t}\PY{p}{)}, \PY{n+nv}{postps} \PY{p}{(}\PY{n+nf}{postp} \PY{n+nv}{t}\PY{p}{)}\PY{p}{]}
        \PY{p}{(}\PY{k}{if }\PY{p}{(}\PY{n+nb}{= }\PY{l+m+mi}{1} \PY{p}{(}\PY{n+nb}{count }\PY{n+nv}{preps}\PY{p}{)} \PY{p}{(}\PY{n+nb}{count }\PY{n+nv}{postps}\PY{p}{)}\PY{p}{)}
          \PY{p}{(}\PY{k}{let }\PY{p}{[}\PY{n+nv}{q} \PY{p}{(}\PY{n+nb}{first }\PY{n+nv}{preps}\PY{p}{)}, \PY{n+nv}{r} \PY{p}{(}\PY{n+nb}{first }\PY{n+nv}{postps}\PY{p}{)}\PY{p}{]}
            \PY{p}{(}\PY{k}{if }\PY{p}{(}\PY{n+nb}{or }\PY{p}{(}\PY{n+nb}{identical? }\PY{n+nv}{q} \PY{n+nv}{r}\PY{p}{)}
                    \PY{p}{(}\PY{n+nb}{and }\PY{p}{(}\PY{n+nb}{not }\PY{p}{(}\PY{n+nf}{member?} \PY{n+nv}{r} \PY{p}{(}\PY{n+nf}{adjs} \PY{n+nv}{q} \PY{l+s+ss}{:pret} \PY{l+s+ss}{:postp}\PY{p}{)}\PY{p}{)}\PY{p}{)}
                         \PY{p}{(}\PY{n+nb}{not }\PY{p}{(}\PY{n+nf}{member?} \PY{n+nv}{r} \PY{p}{(}\PY{n+nf}{adjs} \PY{n+nv}{q} \PY{l+s+ss}{:postt} \PY{l+s+ss}{:prep}\PY{p}{)}\PY{p}{)}\PY{p}{)}\PY{p}{)}\PY{p}{)}
              \PY{p}{(}\PY{k}{let }\PY{p}{[}\PY{n+nv}{merger} \PY{p}{(}\PY{o}{@}\PY{n+nv}{place2or} \PY{n+nv}{q}\PY{p}{)}, \PY{n+nv}{mergee} \PY{p}{(}\PY{o}{@}\PY{n+nv}{place2or} \PY{n+nv}{r}\PY{p}{)}\PY{p}{]}
                \PY{p}{(}\PY{n+nb}{when\PYZhy{}not }\PY{p}{(}\PY{n+nb}{identical? }\PY{n+nv}{q} \PY{n+nv}{r}\PY{p}{)}
                  \PY{p}{(}\PY{n+nf}{eaddall!} \PY{n+nv}{q} \PY{l+s+ss}{:pret}  \PY{p}{(}\PY{n+nf}{eget\PYZhy{}raw} \PY{n+nv}{r} \PY{l+s+ss}{:pret}\PY{p}{)}\PY{p}{)}
                  \PY{p}{(}\PY{n+nf}{eaddall!} \PY{n+nv}{q} \PY{l+s+ss}{:postt} \PY{p}{(}\PY{n+nf}{eget\PYZhy{}raw} \PY{n+nv}{r} \PY{l+s+ss}{:postt}\PY{p}{)}\PY{p}{)}
                  \PY{p}{(}\PY{n+nf}{edelete!} \PY{n+nv}{r}\PY{p}{)}
                  \PY{p}{(}\PY{n+nf}{eaddall!} \PY{n+nv}{merger} \PY{l+s+ss}{:contains} \PY{p}{(}\PY{n+nf}{eget\PYZhy{}raw} \PY{n+nv}{mergee} \PY{l+s+ss}{:contains}\PY{p}{)}\PY{p}{)}
                  \PY{p}{(}\PY{n+nf}{edelete!} \PY{n+nv}{mergee}\PY{p}{)}\PY{p}{)}
                \PY{p}{(}\PY{n+nf}{edelete!} \PY{n+nv}{t}\PY{p}{)}
                \PY{p}{(}\PY{n+nf}{recur} \PY{p}{(}\PY{n+nb}{rest }\PY{n+nv}{ts}\PY{p}{)} \PY{n+nv}{true}\PY{p}{)}\PY{p}{)}
              \PY{p}{(}\PY{n+nf}{recur} \PY{p}{(}\PY{n+nb}{rest }\PY{n+nv}{ts}\PY{p}{)} \PY{n+nv}{applied}\PY{p}{)}\PY{p}{)}\PY{p}{)}
          \PY{p}{(}\PY{n+nf}{recur} \PY{p}{(}\PY{n+nb}{rest }\PY{n+nv}{ts}\PY{p}{)} \PY{n+nv}{applied}\PY{p}{)}\PY{p}{)}\PY{p}{)}
      \PY{n+nv}{applied}\PY{p}{)}\PY{p}{)}\PY{p}{)}

\PY{p}{(}\PY{k+kd}{defn }\PY{n+nv}{assign\PYZhy{}hyperedges} \PY{p}{[}\PY{n+nv}{sc}\PY{p}{]}
  \PY{p}{(}\PY{n+nb}{doseq }\PY{p}{[}\PY{n+nv}{e} \PY{p}{(}\PY{n+nf}{eallobjects} \PY{n+nv}{sc} \PY{l+s+ss}{\PYZsq{}HyperEdge}\PY{p}{)}\PY{p}{]}
    \PY{p}{(}\PY{n+nf}{eset!} \PY{n+nv}{e} \PY{l+s+ss}{:rcontains}
           \PY{p}{(}\PY{n+nb}{first }\PY{p}{(}\PY{n+nb}{apply }\PY{n+nv}{clojure.set/intersection}
                         \PY{p}{(}\PY{n+nb}{map }\PY{o}{\PYZsh{}}\PY{p}{(}\PY{n+nf}{reachables} \PY{n+nv}{\PYZpc{}} \PY{p}{[}\PY{n+nv}{p\PYZhy{}+} \PY{n+nv}{\PYZhy{}\PYZhy{}\PYZlt{}\PYZgt{}}\PY{p}{]}\PY{p}{)}
                              \PY{p}{(}\PY{n+nb}{concat }\PY{p}{(}\PY{n+nf}{eget} \PY{n+nv}{e} \PY{l+s+ss}{:next}\PY{p}{)} \PY{p}{(}\PY{n+nf}{eget} \PY{n+nv}{e} \PY{l+s+ss}{:rnext}\PY{p}{)}\PY{p}{)}\PY{p}{)}\PY{p}{)}\PY{p}{)}\PY{p}{)}\PY{p}{)}\PY{p}{)}

\PY{p}{(}\PY{k+kd}{defn }\PY{n+nv}{create\PYZhy{}top} \PY{p}{[}\PY{n+nv}{sc}\PY{p}{]}
  \PY{p}{(}\PY{k}{let }\PY{p}{[}\PY{n+nv}{top\PYZhy{}ors} \PY{p}{(}\PY{n+nb}{filter }\PY{o}{\PYZsh{}}\PY{p}{(}\PY{n+nb}{not }\PY{p}{(}\PY{n+nf}{eget} \PY{n+nv}{\PYZpc{}} \PY{l+s+ss}{:rcontains}\PY{p}{)}\PY{p}{)} \PY{p}{(}\PY{n+nf}{eallobjects} \PY{n+nv}{sc} \PY{l+s+ss}{\PYZsq{}OR}\PY{p}{)}\PY{p}{)}\PY{p}{]}
    \PY{p}{(}\PY{n+nb}{when }\PY{p}{(}\PY{n+nb}{= }\PY{l+m+mi}{1} \PY{p}{(}\PY{n+nb}{count }\PY{n+nv}{top\PYZhy{}ors}\PY{p}{)}\PY{p}{)}
      \PY{p}{(}\PY{k}{let }\PY{p}{[}\PY{n+nv}{statechart} \PY{p}{(}\PY{n+nf}{ecreate!} \PY{n+nv}{sc} \PY{l+s+ss}{\PYZsq{}Statechart}\PY{p}{)}, \PY{n+nv}{top} \PY{p}{(}\PY{n+nf}{ecreate!} \PY{n+nv}{sc} \PY{l+s+ss}{\PYZsq{}AND}\PY{p}{)}\PY{p}{]}
        \PY{p}{(}\PY{n+nf}{eset!} \PY{n+nv}{statechart} \PY{l+s+ss}{:topState} \PY{n+nv}{top}\PY{p}{)}
        \PY{p}{(}\PY{n+nf}{eset!} \PY{n+nv}{top} \PY{l+s+ss}{:contains} \PY{n+nv}{top\PYZhy{}ors}\PY{p}{)}\PY{p}{)}\PY{p}{)}\PY{p}{)}\PY{p}{)}

\PY{p}{(}\PY{k+kd}{defn }\PY{n+nv}{create\PYZhy{}statechart} \PY{p}{[}\PY{n+nv}{pn}\PY{p}{]}
  \PY{p}{(}\PY{k}{let }\PY{p}{[}\PY{p}{[}\PY{n+nv}{sc} \PY{n+nv}{place2or} \PY{n+nv}{\PYZus{}} \PY{n+nv}{\PYZus{}}\PY{p}{]} \PY{p}{(}\PY{n+nf}{init/init\PYZhy{}statechart} \PY{n+nv}{pn}\PY{p}{)}
        \PY{n+nv}{place2or} \PY{p}{(}\PY{n+nf}{atom} \PY{n+nv}{place2or}\PY{p}{)}\PY{p}{]}
    \PY{p}{(}\PY{n+nf}{iteratively} \PY{p}{(}\PY{k}{fn }\PY{p}{[}\PY{p}{]}
                   \PY{p}{(}\PY{k}{let }\PY{p}{[}\PY{n+nv}{r}     \PY{p}{(}\PY{n+nf}{and\PYZhy{}rule} \PY{n+nv}{pn} \PY{n+nv}{sc} \PY{n+nv}{prep}  \PY{n+nv}{place2or}\PY{p}{)}
                         \PY{n+nv}{r} \PY{p}{(}\PY{n+nb}{or }\PY{p}{(}\PY{n+nf}{and\PYZhy{}rule} \PY{n+nv}{pn} \PY{n+nv}{sc} \PY{n+nv}{postp} \PY{n+nv}{place2or}\PY{p}{)} \PY{n+nv}{r}\PY{p}{)}
                         \PY{n+nv}{r} \PY{p}{(}\PY{n+nb}{or }\PY{p}{(}\PY{n+nf}{or\PYZhy{}rule}  \PY{n+nv}{pn} \PY{n+nv}{sc} \PY{n+nv}{place2or}\PY{p}{)}       \PY{n+nv}{r}\PY{p}{)}\PY{p}{]}
                     \PY{n+nv}{r}\PY{p}{)}\PY{p}{)}\PY{p}{)}
    \PY{p}{(}\PY{n+nf}{create\PYZhy{}top} \PY{n+nv}{sc}\PY{p}{)}
    \PY{p}{(}\PY{n+nf}{assign\PYZhy{}hyperedges} \PY{n+nv}{sc}\PY{p}{)}
    \PY{n+nv}{sc}\PY{p}{)}\PY{p}{)}
\end{Verbatim}